\documentclass[
    ,final            
  ]
  {aipproc}

\layoutstyle{8x11single}

\begin{document}

\title{On a local hidden-variable model with `isolato' hypothesis of the EPR--Bohm Gedanken experiment}




\author{Satoshi Uchiyama}{
  address={Department of Life and Creative Sciences \\
Hokusei Gakuen University Junior College\\
Atsubetsu-ku, Sapporo 004-8631, Japan},
  email={uchiyama@hokusei.ac.jp},
  thanks={The work has been supported in part by the Hokusei Gakuen University Grant for Special Research Projects.}
}

\copyrightyear  {2005}

\begin{abstract}
A local hidden-variable model based on `isolato' hypothesis of the EPR--Bohm Gedanken experiment is presented.
The `isolato' hypothesis states that one of a pair of spin-${1}/{2}$ particles can be in `isolato' mode in which the spin-${1}/{2}$ particle shuts itself from any external interactions, and hence it can never be detected.
As a result of this, data rejection is made; Bell's inequality is violated, although the model is genuinely local.
In this model, $2/\pi$ ($\simeq$ 63.6\%) of the initially prepared ensemble of pairs of spin-${1}/{2}$ particles are detected as a pair of particles, and others are detected as a single particle.
This does not disagree with the results of the experiments performed before, since these single spin-${1}/{2}$ particles were regarded as noise.
\end{abstract}

\date{\today}

\maketitle

\section{Introduction}

The locality principle can be stated from the point of view of a realist as follows: a sharp value for an observable cannot be changed into another sharp value by altering the setting of a remote piece of apparatus~\cite{Redhead}.
From this locality principle and some natural premises, Bell derived the so-called ``Bell's inequality''~\cite{Bell} in the EPR--Bohm Gedanken experiment which is the Bohm version~\cite{Bohm} of the Einstein--Podolsky--Rosen argument~\cite{EinsteinPodolskyRosen}.
In this Gedanken experiment, a pair of spatially separated spin-1/2 particles in a singlet state is considered.
Bell's inequality imposes restrictions on correlations of spins of the particles in the pair.
The statistical prediction of quantum mechanics does not obey this restriction, i.e., the Bell's inequality. 
Many EPR type experiments were performed and violation of Bell type inequalities were reported~\nocite{FreedmanClauser, FryThompson, AspectGrangierRoger:1981, AspectGrangierRoger:1982, AspectDalibardRoger, PerrieDuncanBeyerKleinpoppen}~\cite{FreedmanClauser}--\cite{PerrieDuncanBeyerKleinpoppen}.
Some people have abandoned the locality principle or realism.
Some other people have searched loopholes to preserve the locality principle.
In order to make quantum mechanics compatible with local hidden-variable theory, data rejection hypothesis~\cite{Pearle} which means that some data are rejected by some reason has been investigated by many authors; the lowness of the efficiency of detectors or the procedure of coincidence counting  which cause the data rejection were considered to be loopholes~\nocite{ClauserHorne, MarshallSantosSelleri, Scalera, Notarigo, Pascazio, FerreroMarshallSantos, Uchiyama}~\cite{ClauserHorne}--\cite{Uchiyama}. 
We can find recent attempts to preserve the locality principle in the EPR type experiments in Ref. \cite{VaxjoConf04} and in the references therein.

By data rejection, the probability space of data varies according to contexts of measurement;
we need many probability spaces corresponding to contexts of measurement.
This seems to have a deep connection with the No-Go theorems for contextual hidden-variable theory which says that a quantum-mechanical system cannot be described by a single probability space in general~\cite{Gleason, KochenSpecker}.
In this sense, the cause of the data rejection may possibly be an essential part of quantum mechanics.
Hence it is worth speculating about the meaning of data rejection.
In this paper, we form a new, daring hypothesis that causes the data rejection, and make a local hidden-variable model that reproduces statistical prediction of quantum mechanics exactly.
This model is not so realistic, but simple and genuinely local.
We show that $2/\pi$\% of the initial ensemble of pairs of spin-${1}/{2}$ particles are detected as pairs, and others are detected as single particles.
We also observe that this rate is not varied according to change of context of measurement.

\section{a simple hidden-variable model of the EPR--Bohm Gedanken experiment}

In the EPR--Bohm Gedanken experiment, spin projections of a pair of spin-${1}/{2}$ particles are measured.
We assume that each particle of a pair moves to opposite directions along the $y$-axis.
A direction in a plane parallel to the $zx$-plane is parametrized by an angle $\theta$ from the $z$-direction.
Let $S(\theta)$ be a spin projection to the direction $\theta$.
According to quantum mechanics, $S(\theta)$ assumes only $\pm {\hbar}/{2}$.

Suppose that the set of hidden states of a spin-${1}/{2}$ particle is a circle $C$ which is parametrized by a real parameter $x$ $( -V < x \leq V)$, where $V$ is a positive constant.

We denote a particle that goes along the minus direction of  $y$-axis and a particle goes along the plus direction of $y$-axis by symbols $A$ and $B$, respectively;
hereafter, a suffix $A (/B)$ is added to symbols relating to the particle $A (/B)$.
Then, for a pair of spin-${1}/{2}$ particles $A$ and $B$, the set of hidden states is a torus $T = C_A \times C_B$.
Using the parameters of the circles, $T$ is represented by a set of values of the parameters $(x_A, x_B)$, i.e.
\begin{equation}
T = \left\{ (x_A,\ x_B) \ \left| \  -V < x_A \leq V,\ -V < x_B \leq V \right. \right\}.
\end{equation}

Hereafter, for the sake of simplicity of expressions, the range of the parameters $x_A$ and $x_B$ are extended to the set of all real numbers by identifying $x_A$ with $x_A + 2V$ and $x_B$ with $x_B + 2V$, respectively.

We define the spin projection $S_A(\theta)$ to a direction $\theta$ of the particle $A$ as such a function on $C_A$ that
\begin{equation}
S_A(\theta)(x_A) := \left\{ \begin{array}{ll}
+\frac{\hbar}{2}, & \mbox{ if }  0< x_A - \frac{V}{\pi} \theta \leq V,\\
& \\
-\frac{\hbar}{2}, & \mbox{ if }  -V < x_A - \frac{V}{\pi} \theta \leq 0.\\
\end{array} \right.   \label{SpinProj}
\end{equation}
$S_A(\theta)$ is extended to the function on $T$ in the canonical way.
We define $S_B(\theta)(x_B)$ in the same way.

We now investigate what kind of ensembles of hidden states reproduce the statistical prediction of quantum mechanics in measurements of $S_A(\theta_A)$ and $S_B(\theta_B)$ for a singlet state.
We denote the probability distribution of such an ensemble by $\rho_{(\theta_A, \theta_B)}(x_A, x_B)dx_A dx_B$.
Let $P_{++}(\theta_A, \theta_B)$ be the probability that $S_A(\theta_A) = + {\hbar}/{2}$ and $S_B(\theta_B) = + {\hbar}/{2}$.
Let $P_{--}$ be the probability that $S_A(\theta_A) = - {\hbar}/{2}$ and $S_B(\theta_B) = - {\hbar}/{2}$.
In the same way, let $P_{\pm\mp}$ be the probability that $S_A(\theta_A) = \pm {\hbar}/{2}$ and $S_B(\theta_B) = \mp {\hbar}/{2}$.

Quantum mechanics tells us that 
\begin{eqnarray}
P_{\pm\pm}(\theta_A, \theta_B) &=& \frac{1}{2} \sin^2(\frac{\theta_A - \theta_B}{2}) \\
P_{\pm\mp}(\theta_A, \theta_B) &=& \frac{1}{2} \cos^2(\frac{\theta_A - \theta_B}{2}).
\end{eqnarray}

In the case of $\theta_A = \theta_B$, $P_{\pm\pm}(\theta_A, \theta_B) = 0$ and $P_{\pm\mp}(\theta_A, \theta_B) = {1}/{2}$.
Therefore the support of $\rho_{(\theta_A, \theta_B)}(x_A, x_B)dx_A dx_B$ is contained in the subset $\{ (x_A, x_B) \in T \ | \ x_B = x_A - V \ \mbox{(mod $2V$)}  \}$ of $T$.
Using a non-negative periodic function $\sigma(\cdot; \theta_A, \theta_B)$ of real numbers with period $2V$ which may depend on $\theta_A$ and $\theta_B$, we can write $\rho_{(\theta_A, \theta_B)}(x_A, x_B)dx_A dx_B$ as
\begin{eqnarray}
\rho_{(\theta_A, \theta_B)}(x_A, x_B)dx_A dx_B &=& \sigma(x_A; \theta_A, \theta_B) \delta(x_B - x_A + V) dx_A dx_B \\
&=& \sigma(x_B - V; \theta_A, \theta_B) \delta(x_B - x_A + V) dx_A dx_B,
\end{eqnarray}
where $\delta(x)$ is the Dirac's delta function.

We can see by direct calculations that the following $\sigma_1$ and $\sigma_2$ are the solutions:
\begin{eqnarray}
\sigma_1(x_A; \theta_A, \theta_B) &:=& 
\frac{\pi}{4V} |\sin(\frac{\pi}{V}x_A - \theta_A)|.\\
\sigma_2(x_A; \theta_A, \theta_B) &:=& \frac{\pi}{4V} | \sin(\frac{\pi}{V}x_A - \theta_B) |.
\end{eqnarray}
For example, if $0 < ({V}/{\pi})\theta_A < ({V}/{\pi})\theta_B < V$, then we have only to execute the following integrations:
\begin{equation}
P_{++}(\theta_A, \theta_B) = \int_{\frac{V}{\pi}\theta_A}^{\frac{V}{\pi}\theta_A + V} \! \! \! \! \! \! \!dx_A \ 
\int_{\frac{V}{\pi}\theta_B}^{\frac{V}{\pi}\theta_B + V} \! \! \! \! \! \! \!dx_B \ 
\rho_{(\theta_A, \theta_B)}(x_A, x_B) 
 = \int_{\frac{V}{\pi}\theta_A}^{\frac{V}{\pi}\theta_B} \! \! \! \! \! \! \! d x_A \ \sigma(x_A; \theta_A, \theta_B).
\end{equation}
\begin{equation}
P_{--}(\theta_A, \theta_B) = \int_{\frac{V}{\pi}\theta_A - V}^{\frac{V}{\pi}\theta_A } \! \! \! \! \! \! \!dx_A \ 
\int_{\frac{V}{\pi}\theta_B - V}^{\frac{V}{\pi}\theta_B} \! \! \! \! \! \! \!dx_B \ 
\rho_{(\theta_A, \theta_B)}(x_A, x_B) 
 = \int_{\frac{V}{\pi}\theta_A - V}^{\frac{V}{\pi}\theta_B - V} \! \! \! \! \! \! \! d x_A \ \sigma(x_A; \theta_A, \theta_B).
\end{equation}
\begin{equation}
P_{+-}(\theta_A, \theta_B) = \int_{\frac{V}{\pi}\theta_A}^{\frac{V}{\pi}\theta_A + V} \! \! \! \! \! \! \!dx_A \ 
\int_{\frac{V}{\pi}\theta_B - V}^{\frac{V}{\pi}\theta_B} \! \! \! \! \! \! \!dx_B \ 
\rho_{(\theta_A, \theta_B)}(x_A, x_B) 
 = \int^{\frac{V}{\pi}\theta_A + V}_{\frac{V}{\pi}\theta_B} \! \! \! \! \! \! \! d x_A \ \sigma(x_A; \theta_A, \theta_B).
\end{equation}
\begin{equation}
P_{-+}(\theta_A, \theta_B) = \int_{\frac{V}{\pi}\theta_A - V}^{\frac{V}{\pi}\theta_A } \! \! \! \! \! \! \!dx_A \ 
\int_{\frac{V}{\pi}\theta_B}^{\frac{V}{\pi}\theta_B + V} \! \! \! \! \! \! \!dx_B \ 
\rho_{(\theta_A, \theta_B)}(x_A, x_B) 
 = \int^{\frac{V}{\pi}\theta_A}_{\frac{V}{\pi}\theta_B - V} \! \! \! \! \! \! \! d x_A \ \sigma(x_A; \theta_A, \theta_B). 
\end{equation}

The probability distribution $\rho_{(\theta_A, \theta_B)}(x_A, x_B)dx_A dx_B$ varies according to change of $\theta_A$ or $\theta_B$.
Thus this mathematical model is a kind of contextual hidden-variable model.
It seems to be natural to consider that when the setting  of the apparatus for the particle $A$, i.e. $\theta_A$, is changed, both the hidden state of the particle $A$ and the one of the particle $B$ are changed so as to obey $\rho_{(\theta_A, \theta_B)}(x_A, x_B)dx_A dx_B$.
That is to say, the system is nonlocal.
We can see that this model violates the locality principle.
In fact, for a pair of the particles whose hidden state is $(x_A, x_B)$ such that $\sigma_1(x_A; \theta_A, \theta_B) not= 0$, the particle $B$ in the hidden state $x_B$ vanishes when the setting of the apparatus for the particle $A$ is changed into $\theta'_A := (\pi/V)x_A$.
This kind of nonlocality is too spooky to accept naively.
In the next section, we shall introduce a new hypothesis that enables us to avoid this nonlocality.

A remark is in order concerning this simple model.
Its probability distribution is almost the same as the one proposed by Accardi et al~\cite{AccardiVaxjo01, AccardiVaxjo02, AccardiVaxjo04}.
As explained in ref. \cite{AccardiVaxjo04} the EPR-chameleon dynamical system which produces the probability distribution is local, since the probability distribution is a conditional probability distribution with respect to a context of measurement and the influences of measurements are local.
Then some of particles are ruled out by the conditioning.
We are interested in the particles ruled out; our interpretation of the simple model of this section is slightly different from theirs.

\section{`isolato' hypothesis and violation of Bell's inequality}

The mathematical model presented in the previous section has the following remarkable properties:
$\rho_{(\theta_A, \theta_B)}(x_A, x_B)$ $dx_A$ $dx_B$ depends only on $\theta_A$ if the first solution $\sigma _1$ is adopted, and depends only on $\theta_B$ if the second solution $\sigma _2$ is adopted.

These properties enable us to form a hypothesis that {\it one of a pair of spin-${1}/{2}$ particles can be in a mode in which the spin-${1}/{2}$ particle shuts itself from any external interactions};
therefore it can never be detected in that mode.
We call this mode `isolato' mode, hereafter.
Using this terminology, the hypothesis states that one of a pair of particles can be in `isolato' mode, and the other of the pair is in ordinary mode, i.e, it can be always detected.

For the sake of simplicity, we assume that the particle $A$ can be in `isolato' mode and the particle $B$ is not.
Then we adopt the first solution $\sigma_1$ of $\sigma$ for $\rho_{(\theta_A, \theta_B)}(x_A, x_B)dx_A dx_B$.
We note that $\rho_{(\theta_A, \theta_B)}(x_A, x_B)dx_A dx_B$ represents an ensemble of observed pairs which is different from the ensemble of initially prepared pairs.

The `isolate' hypothesis requires additional hidden-variables to the model in the previous section.
We introduce a hidden variable $\lambda_A$ to the previous model that distinguishes mode of the  particle $A$.
The simplest way of doing this is the following:
we assume that the range of it is $0 \leq \lambda_A \leq {\pi}/{4V}$, and that the particle $A$ in `isolato' mode iff $\lambda_A > ({\pi}/{4V})|\sin(({\pi}/{V})x_A -\theta_A)|$.

We make some remarks with respect to the meaning of this assumption.
Consider a continuous function $s_A(\theta_A)$ defined on $C_A$ by
\begin{equation}
s_A(\theta_A)(x_A) := \frac{\hbar}{2} \sin(\frac{\pi}{V}x_A - \theta_A).
\end{equation}
Since $s_A(\theta_A)$ assumes values between $- \hbar/2$ and $+ \hbar/2$ continuously, it can be said a nonquantized version of the spin-projection defined by eq.~(\ref{SpinProj}). 
This $s_A(\theta_A)$ gives us some intuitive meaning of above assumption.
At $x_A = \pm(V/2) + (V/\pi)\theta_A$, $s_A(\theta_A)$ assumes $\pm\hbar/2$ and the particle $A$ cannot be in `isolato' mode.
At $x_A = (V/\pi)\theta_A$, $(V/\pi)\theta_A + V$, $s_A(\theta_A)$ vanishes.
$(V/\pi)\theta_A$ and  $(V/\pi)\theta_A + V$ are the boundary points between the positive domain of $s_A(\theta_A)$ and the negative domain of $s_A(\theta_A)$.
At these boundary points, the paritcle $A$ is almost always in `isolato' mode.
Thus when the value of the physical quantity $s_A(\theta_A)$ is near the quantized values $\pm\hbar/2$, the particle $A$ is in ordinary mode;
when the value of the physical quantity $s_A(\theta_A)$ is far from the quantized values $\pm\hbar/2$, the particle $A$ is in `isolato' mode, to the contrary.

Let $\lambda_B$ ($0 \leq \lambda_B \leq {\pi}/{4V}$) be an additional hidden parameter for the particle $B$.

The hidden-variable space $T$ is extended to $C_A \times [0, {\pi}/{4V}] \times C_B \times [0, {\pi}/{4V}] =: \Gamma$ to introduce the `isolato' hypothesis.
A hidden state of the system is now described by four parameters ($x_A$, $\lambda_A$, $x_B$, $\lambda_B$).
We assume that the probability distribution $\xi(x_A, \lambda_A, x_B, \lambda_B) dx_A dx_B d\lambda_A d\lambda_B$ that represents the ensemble of the initially prepared pairs is written as
\begin{equation}
\xi(x_A, \lambda_A, x_B, \lambda_B) dx_A dx_B d\lambda_A d\lambda_B = \frac{16 V^2}{\pi^2} \delta(x_B - x_A + V) dx_A dx_B d\lambda_A d\lambda_B.
\end{equation}
In this ensemble, hidden states $x_A$ and $x_B$ are distributed uniformly over the circles $C_A$ and $C_B$, respectively.
The initially prepared ensemble does not depend on the choice of settings $(\theta_A, \theta_B)$ of the apparatuses.
To proceed, we consider an unnormalized distribution $\Xi dx_A dx_B d\lambda_A d\lambda_B$, instead of the normalized one $\xi dx_A dx_B d\lambda_A d\lambda_B$, as a representation of the initially prepared ensemble:
\begin{equation}
\Xi(x_A, \lambda_A, x_B, \lambda_B) dx_A dx_B d\lambda_A d\lambda_B := \frac{4 V}{\pi} \delta(x_B - x_A + V) dx_A dx_B d\lambda_A d\lambda_B.
\end{equation}
By virtue of the `isolato' hypothesis, the probability distribution  of observed pairs which we denote by $\eta_{(\theta_A, \theta_B)} dx_A d\lambda_A dx_B d\lambda_B$ becomes 
\begin{equation}
\eta_{(\theta_A, \theta_B)}(x_A, \lambda_A, x_B, \lambda_B) dx_A d\lambda_A dx_B d\lambda_B = \frac{4V}{\pi}\chi_{\frac{\pi}{4V}|\sin(\frac{\pi}{V}x_A -\theta_A)|}(\lambda_A)d\lambda_A d\lambda_B \delta(x_B - x_A +V)dx_A dx_B,
\end{equation}
where we use a symbol $\chi_X$ to represent the characteristic function of a set $X$.
Note that $\eta_{(\theta_A, \theta_B)}(x_A, \lambda_A, x_B, \lambda_B)$ $\leq$ $\Xi(x_A, \lambda_A, x_B, \lambda_B)$.
This means that $\eta_{(\theta_A, \theta_B)}dx_A dx_B d\lambda_A d\lambda_B$ represents a subensemble of the initially prepared ensemble represented by $\Xi dx_A dx_B d\lambda_A d\lambda_B$.

For some setting of the apparatus for the particle $A$, the mode of the particle $A$ becomes `isolato' mode, and the particle is not observed.
This change of mode of the particle $A$ happens locally.
The hidden state of the particle $A$ is not affected by a setting of the apparatus for the particle $B$.
Since the particle $B$ is not influenced by a setting of the apparatus for the particle $A$, if the particle $A$ is in `isolato' mode, then the particle $B$ is observed as a single particle.
Such single particles are usually regarded as noise in experiments; they are not taken into account as true data even if they are observed.

It is easy to see that
\begin{equation}
\int_{0}^{\frac{\pi}{4V}} d\lambda_A \int_{0}^{\frac{\pi}{4V}} d\lambda_B \eta_{(\theta_A, \theta_B)}(x_A, \lambda_A, x_B, \lambda_B)  dx_A  dx_B  = \rho_{(\theta_A, \theta_B)}(x_A, x_B) dx_A dx_B.
\end{equation}
Therefore our local hidden-variable model based on the `isolato' hypothesis reproduces exactly the statistical prediction of quantum mechanics except that single particles, the number of which does not depend on setting $(\theta_A, \theta_B)$ of the apparatuses, are always observed.

The `isolato' hypothesis causes the experimenter to reject data of single particles.
It has been known for a long time, that data rejection  makes it possible that a local hidden-variable model violates the Bell's inequality.
The data rejection hypothesis, however, is not taken seriously by the majority, since, as the author of Ref. ~\cite{Pearle} himself says, ``...; had such large fractions of undetected events occurred in other already performed correlation experiments, it is hard to see how such behavior would have gone unnoticed.''
An interpretation based on the `isolato' hypothesis is that single particle data are rejected  by the experimenter who notices their existence.
We have to know how many data are rejected in our model.
By integrating the unnormalized distribution $\Xi dx_A dx_B d\lambda_A d\lambda_B$ over the space of hidden states, we obtain the number of the initially prepared pairs in units such that the number of observed pairs is unity.
\begin{equation}
\int_{-V}^{V}dx_A \int_{-V}^{V}dx_B \int_{0}^{\frac{\pi}{4V}}d\lambda_A  \int_{0}^{\frac{\pi}{4V}}d\lambda_B\Xi(x_A, \lambda_A, x_B, \lambda_B)
= \frac{\pi}{4V} \int_{-V}^{V} dx_A
= \frac{\pi}{2}.
\end{equation}
Therefore $1/ ({\pi}/{2}) = {2}/{\pi} \simeq 63.6\% $ of the initially prepared pairs are observed as pairs in our model.
In other words, single particles are observed ${(\pi -2)}/{2}\simeq 0.571$ times as many as the number of observed pairs.

There are experiments which verified violation of Bell type inequalities~\nocite{FreedmanClauser, FryThompson, AspectGrangierRoger:1981, AspectGrangierRoger:1982, AspectDalibardRoger, PerrieDuncanBeyerKleinpoppen}~\cite{FreedmanClauser}--\cite{PerrieDuncanBeyerKleinpoppen}.
In some of them, singles count rates are reported.
In the experiment performed by Aspect et al.~\cite{AspectGrangierRoger:1981} using photons, the typical singles count rate is 40 000 counts per second and the coincidence count rate is 150 counts per second.
In the experiment by Aspect et al.~\cite{AspectGrangierRoger:1982} using two-channel polarizers, the typical singles count rate is over $10^{4}$ counts per second and the sum of the four coincidence count rates is typically 80 counts per second.
In the experiment performed by Perrie et al.~\cite{PerrieDuncanBeyerKleinpoppen} using photons, the typical singles count rates is about $10^4$ counts per second and the true two-photon coincidence count rate is 490 counts per hour $\simeq$  $10^{-1}$ counts per second.
They estimates that about 0.01\% of the singles count rates is due to uncorrelated photons from two-photon decay process itself and the other is due mainly to radiation produced by interaction of the atomic beam with background gas.
Then the counting rate of singles that cannot be regarded as background noise is about $10^0$ counts per second.
Thus our local hidden-variable model with the `isolato' hypothesis is not falsifed immediately by these experiments.

\section{discussion}

Our local hidden-variable model is restrictive to the typical EPR--Bohm Gedanken experiment.
But it is possible to extend the model by assuming  evolution of hidden states with time so as to describe such a case that the setting $(\theta_A, \theta_B)$ of the apparatuses are changed with time like the experiment performed by Aspect et al.~\cite{AspectDalibardRoger}.

When the pair is produced at time $t_0$, the probability distribution is described by $\xi(x_A, \lambda_A, x_B, \lambda_B) dx_A dx_B d\lambda_A d\lambda_B$ as already stated.
We assume that after the separation of the particles, the hidden state  $(x_A, \lambda_A)$ of the particle $A$ keeps on changing rapidly with evolution of time.
More precisely, we assume that there exists an increasing function $\varphi_A(t)$ of time $t$ such that the particle $A$ is in `isolato' mode iff its hidden variables satisfy $\lambda_A > ({\pi}/{4V})|\sin(({\pi}/{V})x_A -\varphi_A(t))|$.
Note that $\varphi_A(t)$ does not represent a setting of the apparatus for the particle $A$ at this point.
Suppose the particle $A$ reaches the apparatus at time $t_1$.
Then there exists a time $t_2$ such that $\varphi_A(t)$ coincides to the setting parameter $\theta_A$ and $t_2 \geq t_1$.
The second assumption for $\varphi_A(t)$ is that the evolution of $\varphi_A(t)$ stops after $t_2$, i.e. $\varphi_A(t) = \theta_A$ for $t > t_2$.
Then the probability distribution of observed pairs becomes $\eta_{(\theta_A, \theta_B)}(x_A, \lambda_A, x_B, \lambda_B) dx_A d\lambda_A dx_B d\lambda_B$ as before.
It is clear that these assumptions on evolution of the hidden state are compatible with the locality principle.

It is not difficult to make the initially prepared ensemble symmetric with respect to the exchange of the particle $A$ and the particle $B$  without violating the locality principle.
We introduce a two-valued parameter $\mu$ such that if $ -1 \leq \mu < 0$, then the particle $A$ may be in `isolato' mode and the particle $B$ is in ordinary mode; if $0 \leq \mu \leq 1$, then the particle $A$ is in ordinary mode and the particle $B$ may be in `isolate' mode.
$\mu$ is an additional hidden-variable for the pair.
We assume that the probability that $-1 \leq \mu < 0$ is equal to the one that $ 0 \leq \mu \leq 1$.
We can consider that the value of $\mu$ is determined when the two particles in a pair separate from each other.
Hence introduction of $\mu$ does not violate the locality principle.
The condition of being in `isolato' mode has to be changed slightly;
the particle $A$ in `isolato' mode iff $\lambda > ({\pi}/{4V})|\sin(({\pi}/{V})x_A -\theta_A)|$ and $\mu \in [-1, 0)$;
the particle $B$ in `isolato' mode iff $\lambda > ({\pi}/{4V})|\sin(({\pi}/{V})x_B -\pi  -\theta_B)|$ and $\mu \in [0, 1]$.

Thus our hidden-variable space $\Gamma$ is extended to $C_A \times [0, {\pi}/{4V}] \times C_B \times [0, {\pi}/{4V}] \times [-1, 1]$.
A hidden state is described by five parameters ($x_A$, $\lambda_A$, $x_B$, $\lambda_B$, $\mu$).
Suppose that the joint probability distribution $\eta_{(\theta_A, \theta_B)}(x_A, \lambda_A, x_B, \lambda_B, \mu) dx_A d\lambda_A dx_B d\lambda_B d\mu$ of these five parameters is given by
\begin{eqnarray*}
\eta_{(\theta_A, \theta_B)} dx_A d\lambda_A dx_B d\lambda_B d\mu &:=& \frac{2V}{\pi}\chi_{\frac{\pi}{4V}|\sin(\frac{\pi}{V}x_A -\theta_A)|}(\lambda_A)d\lambda_A d\lambda_B \delta(x_B - x_A +V)dx_A dx_B \chi_{[-1, 0)}(\mu) d \mu \\
&&+  \frac{2V}{\pi}\chi_{\frac{\pi}{4V}|\sin(\frac{\pi}{V}x_B -\pi -\theta_B)|}(\lambda_B)d \lambda_B d\lambda_A \delta(x_B - x_A +V)dx_A dx_B \chi_{[0, 1]}(\mu) d \mu.
\end{eqnarray*}
Integrating this with respect to the hidden variables $\lambda_A$, $\lambda_B$ and $\mu$, we have
\begin{eqnarray*}
\int_{0}^{\frac{\pi}{4V}} d\lambda_A \int_{0}^{\frac{\pi}{4V}} d\lambda_B \int_{-1}^{1} d\mu \eta_{(\theta_A, \theta_B)} dx_A dx_B  &=& \frac{1}{2} \frac{\pi}{4V}|\sin(\frac{\pi}{V}x_A -\theta_A)|  \delta(x_B - x_A +V)dx_A dx_B  \\
&&+  \frac{1}{2}\frac{\pi}{4V}|\sin(\frac{\pi}{V}x_B -\pi -\theta_B)|  \delta(x_B - x_A +V)dx_A dx_B \\
 &=& \frac{1}{2} \sigma_{1}(x_A)  \delta(x_B - x_A +V)dx_A dx_B  \\
&&+  \frac{1}{2}\sigma_{2}(x_B - V) \delta(x_B - x_A +V)dx_A dx_B .
\end{eqnarray*}
Again, we obtain a probability distribution which reproduces the statistical prediction of quantum mechanics in the EPR--Bohm situation.
The initially prepared ensemble is represented by
\begin{eqnarray*}
\Xi(x_A, \lambda_A, x_B, \lambda_B, \mu) dx_A d\lambda_A dx_B d\lambda_B d\mu &:=& \frac{2V}{\pi} d\lambda_A d\lambda_B \delta(x_B - x_A +V)dx_A dx_B \chi_{[-1, 0)}(\mu) d \mu \\
&&+  \frac{2V}{\pi} d \lambda_B d\lambda_A \delta(x_B - x_A +V)dx_A dx_B \chi_{[0, 1]}(\mu) d \mu.
\end{eqnarray*}
The total number of the initially prepared pairs is obtained by the integration over the hidden variable space $\Gamma$, i.e.
\begin{equation}
\int_{\gamma} \Xi dx_A d\lambda_A dx_B d\lambda_B d\mu = \frac{2V}{\pi} \frac{\pi^2}{16V^2} 2V  + \frac{2V}{\pi} \frac{\pi^2}{16 V^2} 2V = \frac{\pi}{2}.
\end{equation}
Therefore the ratio of the number of observed pairs to the one of the initially prepared pairs is ${2}/{\pi}$ again.

It sounds strange that there exist two kinds of spin-1/2 particle.
A spin-1/2 particle of the first kind is in ordinary mode, and the one of the second kind can be in `isolato' mode.
This could be understood naturally if we consider in the following way.
In order that two spin-1/2 particles form a pair in a singlet state, the third matter which can not be broken into smaller part would be necessary.
The two spin-1/2 particles are joined by the agency of this third matter.
When the two spin-1/2 particles are separated, the third matter sticks to one of them, the two spin-1/2 particles become different kinds of spin-1/2 particle.

The `isolato' hypothesis, which says that a particle in `isolato' mode is not observed though it exists, may have something to do with interference phenomena, too.
Guessing apart, our local hidden-variable model based on `isolato' hypothesis is refutable experimentally by removing noise carefully.
It is true that the model is not sufficiently realistic.
It is, however, possible to hope that the model describes some elements of reality in the EPR--Bohm Gedanken experiment, and hence the physical reality obeys the locality principle, until it is refuted by experiments.

\begin{theacknowledgments}
The author would like to thank Prof. A. Yu. Khrennikov for giving me an opportunity of presenting my idea at the stimulative conference, Prof. A. F. Kracklauer for helpful comments and Prof. L. Accardi for useful discussion.
The work has been supported in part by the Hokusei Gakuen University Grant for Special Research Projects. 
\end{theacknowledgments}



\bibliographystyle{aipproc}   

\bibliography{myrefs}

\end{document}